\documentclass[twocolumn,amsmath,amssymb,superscriptaddress]{revtex4}
\usepackage[latin9]{inputenc}
\usepackage{amsmath}
\usepackage{amssymb}
\usepackage{graphicx}
\PassOptionsToPackage{normalem}{ulem}
\usepackage{ulem}

\makeatletter
\@ifundefined{textcolor}{}
{%
 \definecolor{BLACK}{gray}{0}
 \definecolor{WHITE}{gray}{1}
 \definecolor{RED}{rgb}{1,0,0}
 \definecolor{GREEN}{rgb}{0,1,0}
 \definecolor{BLUE}{rgb}{0,0,1}
 \definecolor{CYAN}{cmyk}{1,0,0,0}
 \definecolor{MAGENTA}{cmyk}{0,1,0,0}
 \definecolor{YELLOW}{cmyk}{0,0,1,0}
}

\usepackage{float}
\usepackage{dcolumn}
\usepackage{bm}

\newcommand{\er}[1]{(\ref{#1})}
\begin{document}
\def\ts#1{\mbox{$\textsubscript{#1}$}}
\def\ih{\frac{i}{\hbar}}
\def\be{\begin{equation}}
\def\ee{\end{equation}}
\def\bea{\begin{eqnarray}}
\def\eea{\end{eqnarray}}
\def\ds{\displaystyle}
\def\bra#1{\mbox{$\langle#1|$}}
\def\ket#1{\mbox{$|#1\rangle$}}
\def\braket#1{\mbox{$<#1>$}}
\def\midmid#1{\mbox{$|#1|$}}
\def\pd{\displaystyle\frac{\partial}{\partial t}}
\def\rso{{\hat \rho}}

\title{Site--dependence of van der Waals interaction explains exciton spectra
of double--walled tubular J--aggregates}


\author{Jörg Megow}

\email{megow@uni-potsdam.de}


\affiliation{Institut für Chemie, Universität Potsdam, Karl-Liebknecht-Straße
24-25, D-14476 Potsdam, F. R. Germany}

\author{Merle I. S. Röhr}

\affiliation{Institut für Physikalische und Theoretische Chemie, Universität Würzburg,
Emil-Fischer-Straße 42, D-97074 Würzburg, F. R. Germany}

\author{Marcel Schmidt am Busch}

\affiliation{Institut für theoretische Physik, Johannes Kepler Universität Linz,
Altenberger Straße 69, AT-4040 Linz, Austria}

\author{Thomas Renger}


\affiliation{Institut für theoretische Physik, Johannes Kepler Universität Linz,
Altenberger Straße 69, AT-4040 Linz, Austria}

\author{Roland Mitri\'{c}}

\affiliation{Institut für Physikalische und Theoretische Chemie, Universität Würzburg,
Emil-Fischer-Straße 42, D-97074 Würzburg, F. R. Germany}

\author{Stefan Kirstein}


\affiliation{Institut für Physik, Humboldt--Universität zu Berlin, Newtonstraße
15, D-12489 Berlin, F. R. Germany}

\author{Jürgen P. Rabe}


\affiliation{Institut für Physik, Humboldt--Universität zu Berlin, Newtonstraße
15, D-12489 Berlin, F. R. Germany}

\affiliation{IRIS Adlershof, Humboldt--Universität zu Berlin, Zum Großen Windkanal
6, D-12489 Berlin, F. R. Germany}

\author{Volkhard May}


\affiliation{Institut für Physik, Humboldt--Universität zu Berlin, Newtonstraße
15, D-12489 Berlin, F. R. Germany}

\begin{abstract}
The simulation of the optical properties of supramolecular
aggregates requires the development of methods, which are able to
treat a large number of coupled chromophores interacting with
the environment. Since it is currently not possible to treat large systems by quantum chemistry,
the Frenkel exciton model is a valuable alternative. In this work
we show how the Frenkel exciton model can be extended in order to
explain the excitonic spectra of a specific double--walled tubular dye aggregate explicitly taking
into account dispersive energy shifts of ground and excited states
due to van der Waals interaction with all surrounding molecules.
The experimentally
observed splitting is well explained by the site--dependent energy
shift of molecules placed at the inner or outer side of the double--walled tube, respectively.
 Therefore we can conclude,
that inclusion of the site--dependent dispersive effect in the theoretical
description of optical properties of nanoscaled dye
aggregates is mandatory.
\end{abstract}
\maketitle
The investigation of electronic exciton energy transfer (EET)
in huge supramolecular systems is of enormous interest for the understanding
of processes taking place in both biological \cite{Zhang2014, Schulten2011,Scholes2014} and artificial \cite{Eisele2009, Eisele2012}
light harvesting antennae systems and therefore topic of
recent research.
As an example chlorosomes
from green sulfur bacteria which cover thousands of pigment molecules
have been investigated only recently \cite{aspuru12,korppi13,aspuru14}. 

Various artificial light harvesting systems have 
been established and explored experimentally \cite{Wuerthner2011,Lin2010}
and were also investigated theoretically, such as giant molecular
macrocyles \cite{lupton13,lupton13a} or large complexes of tetrapyrrole
type molecules \cite{Ermilov2005,Megow2014}. However, a appropriate
theoretical treatment of such systems still remains challenging.
In this work we focus on the proper description ot the structural
and optical properties of the tubular dye aggregate (TDA) of the
amphiphilic cyanine dye named {\textbf C8S3} \cite{C8S3}, formed in aqueous
solution, which have been discussed in several studies \citep{Berlepsch2007,Eisele2009,Clark2013,Berlepsch2011}.

Former cryogenic transmission electron microscope (cryo-TEM)
images suggested that the dyes are arranged within a double layer
structure similar to lipid membranes facing the hydrophilic groups,
and hence the chromophoric units, towards the liquid medium inside
and outside of the tube and hiding the hydrophobic alkyl chains from
the aqueous environment \citep{Berlepsch2007,Eisele2009}. This building
principle is illustrated in Figures 1 and 2.

In previous studies investigating the double--walled tubular
J-aggregates the linear optical spectra were explained theoretically by a parameterized
Frenkel exciton Hamiltonian \citep{Didraga2004, Pugzlys2006,
Vlaming2011}. Therefore, the tubes were modelled by two telescoped tubes of dye
molecules. For each cylinder independently a regular and destortion
free lattice of dyes was assumed with two molecules per unit cell
where every dye was represented by an extended dipole \cite{Eisele2012,Clark2013,Eisele2014}.
All energy shifts in the spectra were attributed to excitonic coupling
while solvent shifts or dispersive energies were neglected or simulated
by a constant shift. By the adjustment of mutual orientations, distances,
and tilt angles, the absorption spectra were modelled in great detail.
Also an assigment of bands that correspond to the inner and outer
cylinder were provided. Notice, that in those works the structure
of inner and outer cylinder was independent.

In contrast, highly sophisticated image reconstruction techniques
revealed details of the structure of the tube wall indicating
an organization of the dyes in ribbons
that are helically winding within the wall \cite{Berlepsch2011}.
The appearence of these ribbons requires a strong correlation between
the arrangement of dyes within the inner and outer tube, which is
not fulfilled by the model described above.
On the other hand, the computation of optical spectra based on molecular dynamics (MD) simulations was
approached in \cite{Haverkort2014} for a single--walled tube and resulted in a too wide absorption line shape.

\begin{figure}[h]
\includegraphics[width=0.45\textwidth]{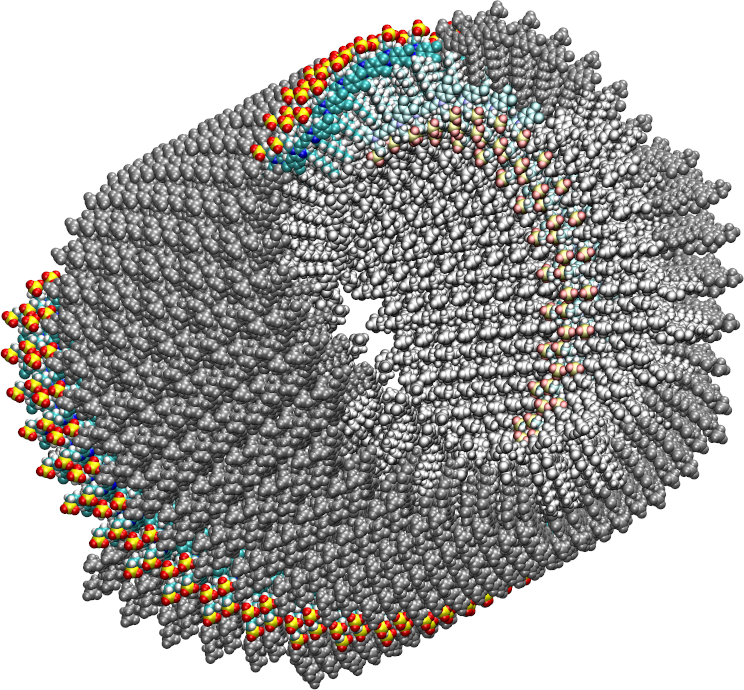}
\caption{{\footnotesize {Initial model of {\textbf C8S3} TDA. One molecular ribbon of the outer
cylinder is highlighted in color. The whole outer cylinder is drawn
in grey and the inner one in light grey. (The figure was created using
VMD \cite{Humphrey1996}.)}}}

{\footnotesize \label{fig1} }
\end{figure}

In this contribution we provide two new approaches to describe the
corrected aggregate structure and spectra: first, the structure is
modelled employing
MD simulations. The start configuration is resolved to best approximate
the appearence of highly resolved cryo-TEM images (cf. Fig. \ref{fig1}).
A detailled model down to atomic resolution has been obtained that
can be handled numerically and is stable over timescales of nanoseconds.
Second, the linear absorption spectrum is calculated based on the
MD structure taking into account transition densities for resonant
excitonic interactions but also a full description of dispersive energies
due to non-resonant interactions. The huge number of molecules within
the slightly disordered aggregates account for disorder effects in
the spectra. The results lead us to a new interpretation of the shift,
and splitting into two fundamental bands of the absorption spectrum.


Considering bulk systems, the change of the molecular excitation energy
due to an environmental influence is known as the gas--to--crystal
shift. Electrostatic and inductive site energy shifts depend on the
difference in charge density between the excited and the ground state
of the molecule. If the latter is small, the leading contribution
to the site energy shift is due to mutual polarization of different
atoms (molecules): the van der Waals interaction or dispersive interaction.
It was first proposed by London \cite{London1937}, to explain the
attractive interaction between noble gas atoms (exhibiting closed
electronic shells). Due to the Coulomb coupling between electrons
in different atoms (molecules) a dynamic correlation in electronic
motion stabilizes the coupled system and leads to reduction in energy.
Since excited state wavefunctions are often more extended than the
ground state ones, the dispersive shifts of excited state energies
are usually larger than those of electronic ground states leading
to a redshift of the transition energy \cite{Renger2008}. It is important
to note that the dispersive shift also includes the interaction with
all other molecules of the environment.

Usually, dispersion effects of the surrounding medium are taken into
account by assuming a single value for the molecular excitation (site)
energy in order to place the whole absorption spectrum in the correct
wavelength region. This implies the use of a constant transition energy
shift. In the case of a nano--structured system the polarizability
of the environment may strongly depend on the molecular position.
Accordingly, the excitation energy shift does not remain a constant
but becomes site--dependent. If this shift overcomes the strength
of the resonant EET (excitonic) coupling it affects significantly
the spectrum of excitation energies. We note, that such site--dependent
dispersive energy shifts have been calculated for aggregates of Rydberg
atoms in \cite{Zoubi2014}. Here we utilize a method that is applicable
to molecular aggregates.


It will be shown in the following that the separation of the two low--energy
absorption bands as reported in \cite{Eisele2012} is mostly due to
different dispersive shifts of the molecular excitation energy in
the inner and the outer wall of the TDA. On the other hand, the resonant
excitonic coupling is found to have less impact on the separation
of these two bands.

\section{Results}

\begin{figure}[h]
\includegraphics[width=0.25\textwidth]{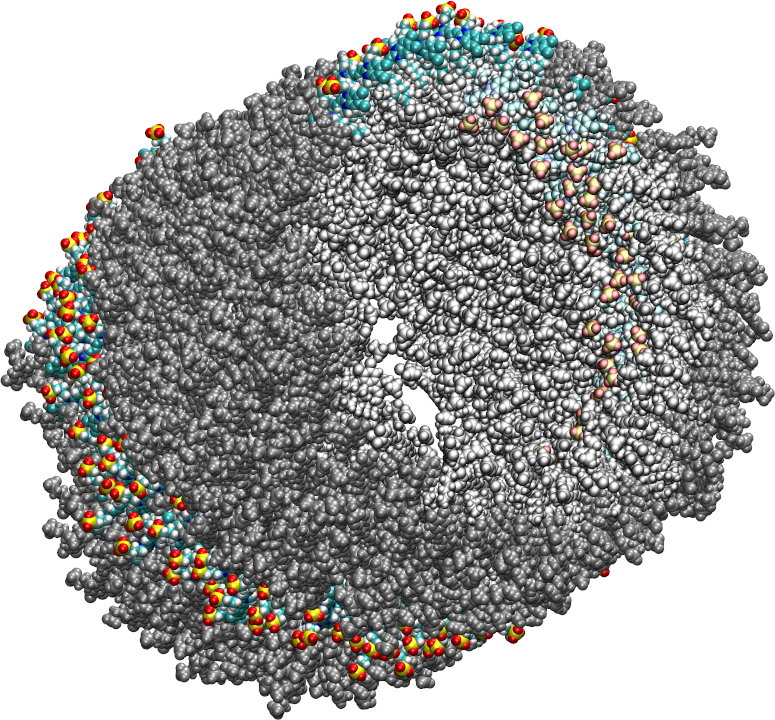}
\includegraphics[width=0.22\textwidth]{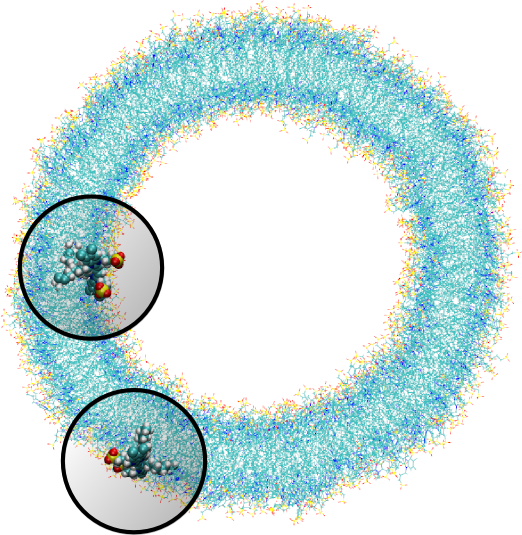}
\caption{{\footnotesize { MD aggregate after 7 ns of simulation.
Upper panel:
side view according to Fig.~\ref{fig1}. Lower panel: 
Top view of
the {\textbf C8S3} TDA with a dye molecule in the inner cylinder and the outer
cyliner highlighted. The spheres around the highlighted molecules
have a radius of 2 nm. They indicate a volume that includes all dye
molecules that contribute more than 99\% of the contributions to the
dispersive energy shift. Note that more dye molecules are within this
sphere for the molecules at the inner wall then those for the outer
wall. (The figures was created using VMD \cite{Humphrey1996}.)}}}
{\footnotesize \label{fig2} }
\end{figure}

To achieve a consistent explanation of the TDA absorption spectrum,
our subsequent considerations will be based on a spatial structure
that is obtained by MD simulations starting from a configuration resembling
highly resolved cryo-TEM images \cite{Berlepsch2011}. The simulated
system included 828 {\textbf C8S3} molecules, 74116 water molecules and 9242
methanol molecules in a box of 17 $\times$ 17 $\times$ 12.6 nm with
periodic boundary conditions. 
The simulations are extended up to 7
ns. Here, it is essential to ensure that the
solvent density inside the tube coincides with that outside. After
an equilibration time of a few 100 ps the TDA structural parameters
were found to stay constant, indicating a stable structure of the
aggregate. In Fig.~\ref{fig2} the structure after 7 ns of simulation
is shown.

In the case of a TDA of {\textbf C8S3}, due to the curvature the environment
in proximity of the dye molecules has a different composition in terms
of number of dye molecules per volume. This mainly depends on the
molecule either being located in the outer wall or in the inner wall.
This effect is indicated in the lower panel of Fig.~\ref{fig2} by
the circles showing the close environment for two dyes within the
top view of a MD simulated aggregate. This circles with a radius of
about 2 nm around the center of the molecular transition dipole moments
indicate the close environment of the dyes which contributes most
($>$99\%) to the dispersive energy shift.

\begin{figure}[h!]
\includegraphics[width=0.29\textwidth]{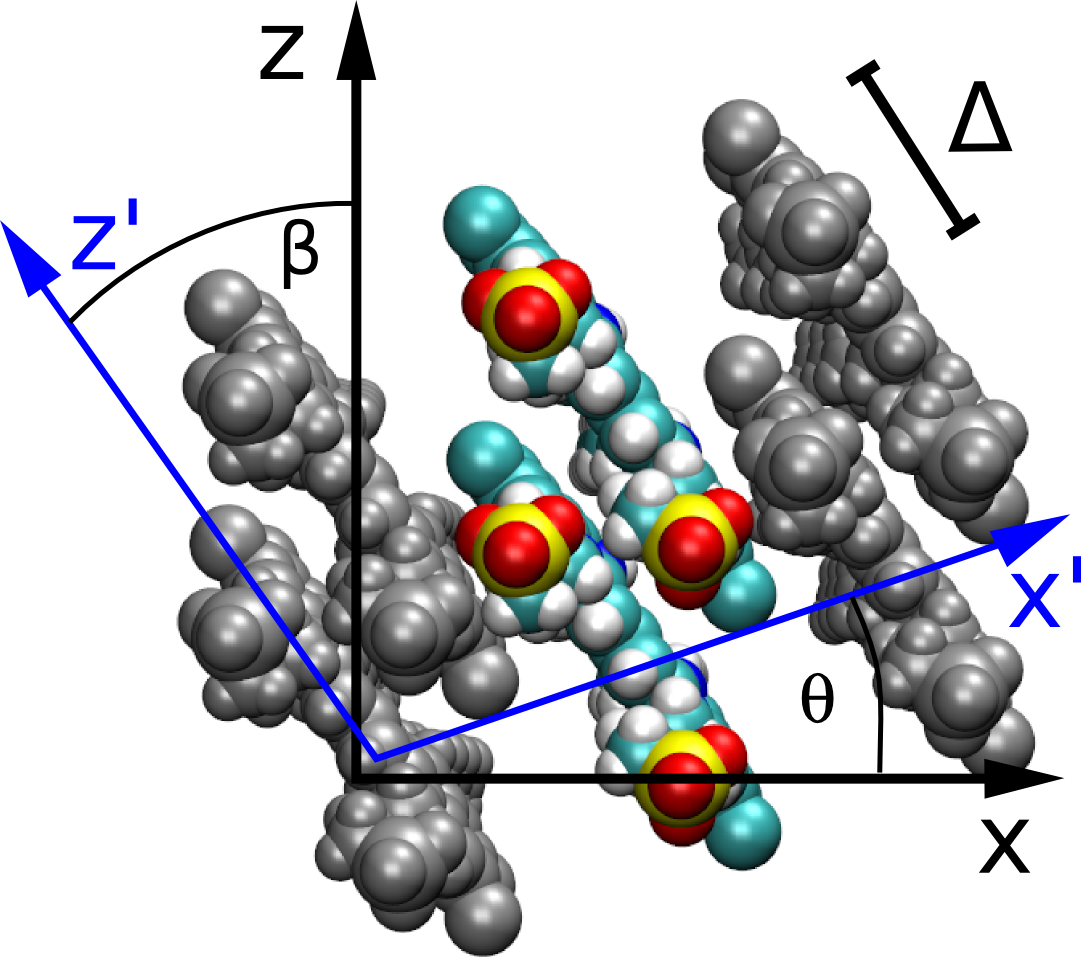}
\includegraphics[width=0.18\textwidth]{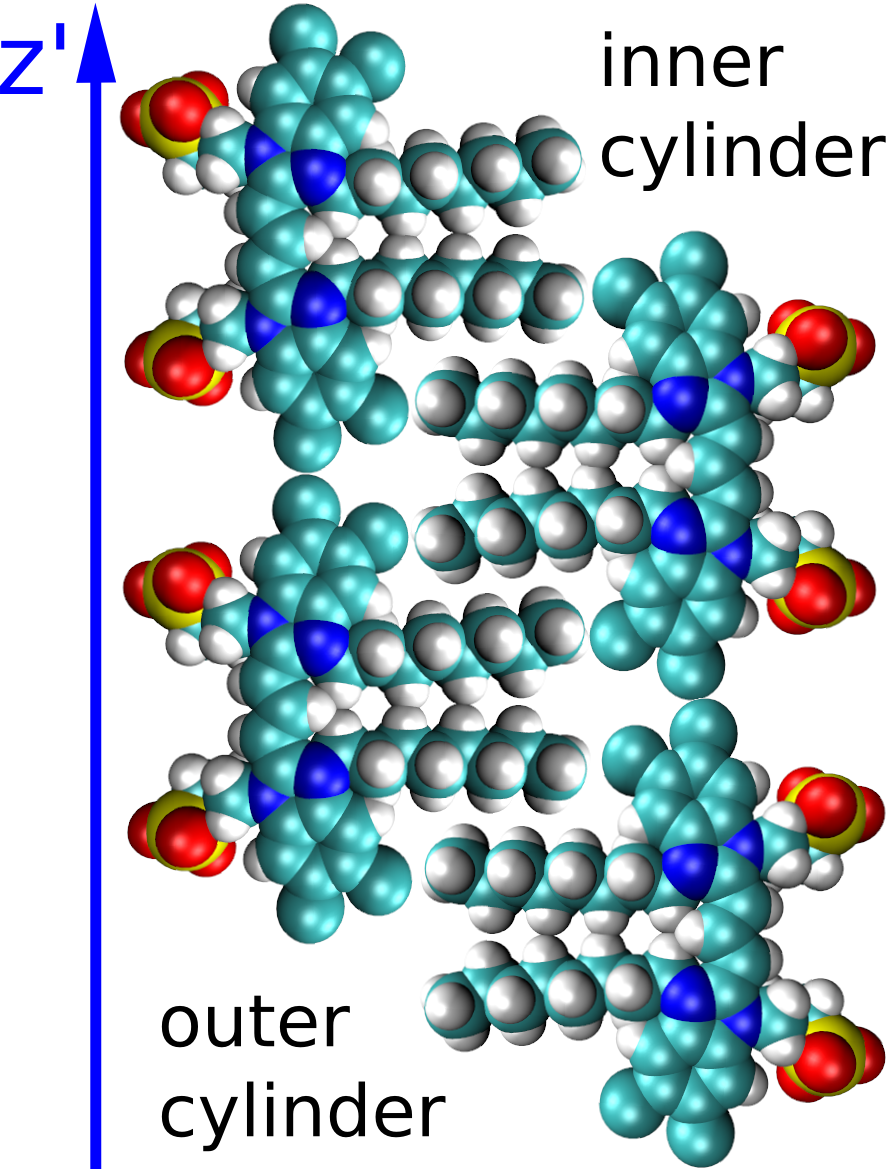}
\caption{{\footnotesize {Internal structure of the {\textbf C8S3} TDA. Left panel: molecular
ribbon with two molecules forming a unit cell. The internal coordinate
system is drawn in blue. Right panel (schematic): orientation of the
{\textbf C8S3} dyes in the outer (left) and inner (right) cylinder. (The figures
were created using VMD \cite{Humphrey1996}.)}}}

{\footnotesize \label{fig3} }
\end{figure}

\subsection{TDA structure}
According to the cryo--TEM data, the whole TDA structure consists
of six ribbons, turned around a common axis, forming an inner and
an outer cylinder (see Fig. \ref{fig1}). Both cylinders have the
same periodicity (twists per aggregate length), which is 12.7 nm.
The width of a single ribbon amounts to 20.1 \AA{}{ } for the outer
cylinder and 19.5 \AA{}{ } for the inner cylinder (they differ due
to the different screw angles). The extension of a single {\textbf C8S3} molecule
along the transition dipole axis is 20.8 \AA{}. We assumed an orientation
of the molecules with dipole moments parallel to the cylinder surface
what is confirmed by our MD simulations. These ribbons then wind up
with a screw angle $\theta$ and the molecules have to be tilted against
the z-axis of the tube by an angle $\beta$ (see Fig. \ref{fig3}).
$\theta$ amounts to 18.3$^{\circ}$ for the outer and 23.0$^{\circ}$
for the inner cylinder. $\beta$ amounts to 34.5$^{\circ}$ for the
outer and 44.3$^{\circ}$ for the inner cylinder. We note that no
exact value for $\beta$ can be extracted from the cryo-TEM images.

The mean distance of neighboring molecules is 5 \AA{}{ } within
both cylinders, measured as the distance between the molecular centers
of the $\pi$-electron system. The closest distance between two molecules
is about 4 \AA{}{ }. The distance between neighbored molecules varies
due to the angle between them. We also tried simulations with closer
packed molecules. They were not stable since some of the atoms came
too close to each other. The distance between the two cylinders amounts
to 13 \AA{}{ } (measured with respect to the centers of the molecular
$\pi$--electron systems).

Our model assumes two molecules per unit cell. Therefore, a mutual
displacement $\Delta$ as indicated in Fig. \ref{fig3} was introduced
by hand with a value of $\Delta=9$ \AA{}. $\Delta$ was chosen in
order to obtain the correct absorption spectrum (see below) and is
the only free parameter in our model. Due to the displacement $\Delta$
our model includes two molecules per unit cell. The structural stability
of the TDA for such a mutual molecular displacement has been confirmed
by extended MD simulations. We note, that using $\Delta=0$, a structure
model consistent with \cite{Berlepsch2011} results in an H--aggregate--like
absorption spectrum.

For comparison, a MD simulation, basing on the structure model presented
in \cite{Eisele2012} was employed. It is specific for that model,
that the molecular structures of the inner and outer wall differ and
furthermore are completely independent from each other, \textit{i.e.}
there is no correlation between the respective molecular positions.
Additionally, the molecules are arranged in a Herringbone-like structure,
also resulting in 2 molecules per unit cell. However, by performing
MD simulations, this model turned out to be unstable, since the tubes
disintegrated with time. Therefore, we conclude that the strong correlation
between inner and outer tube as it is provided by the model of the
two-layer ribbons reduces the system's free energy and stabilizes
the structure significantly.

\subsection{Frenkel exciton Hamiltonian}
Having determined the TDA structure we can turn to a computation of
the resulting excited electronic states. Singly excited states of
the TDA as detected in linear absorption measurements can be deduced
from the standard Frenkel--exciton Hamiltonian
\bea
\label{eq:Hagg}
H_{\rm exc} = \sum_{m} {\cal E}_{m} \ket{\phi_m}\bra{\phi_m} 
+ \sum_{m \neq n} {\cal J}_{m n} \ket{\phi_m}\bra{\phi_n} \; .
\eea
In its diagonal part it covers the so--called site energies
${\cal E}_{m}$ (excitation energy of molecule $m$). The off--diagonal
part is formed by the excitonic coupling ${\cal J}_{mn}$ which resonantly
transfers excitation energy from one dye molecule to the other. The
$\phi_{m}$ are molecular product states with molecule $m$ in the
first excited state $\varphi_{me}$ and all other molecules in the
ground state $\varphi_{ng}$. The restriction to a single excited
state becomes possible since higher excited states are energetically
sufficiently far away. The overall ground state $\phi_{0}$ is a product
of all $\varphi_{ng}$. A ground state part $E_{0}\ket{\phi_{0}}\bra{\phi_{0}}$
does not appear in the Hamiltonian since we set $E_{0}=\sum_{m}E_{mg}=0$.

The excitonic coupling ${\cal J}_{mn}$ coincides with $J_{mn}(eg,eg)$
which is a particular example of the general two--molecule Coulomb
matrix element $J_{mn}(ab,b'a')$. The electronic quantum numbers
$a$ and $a'$ belong to molecule $m$ and $b$ and $b'$ to molecule
$n$. The matrix element appears as the Coulomb interaction of the
charge density $n_{aa'}^{(m)}({\bf x})$ of molecule $m$ and the
charge density $n_{bb'}^{(n)}({\bf y})$ of molecule $n$. If, for
example, $a\neq a'$ the density $n_{aa'}^{(m)}({\bf x})$ is exclusively
determined by the electronic transition density $\rho_{aa'}^{(m)}({\bf x})$
\cite{May2011}. For concrete computations the continuous charge densities
are replaced by discrete atomic centered partial charges and partial
transition charges \cite{Madjet2006}.


\subsection{Curvature induced dispersion shifts}
The minimal version of the exciton Hamiltonian introduced so far has
to be improved by a consideration of further couplings among the molecules
which may change the site energies as well as the matrix elements
of the excitonic coupling \cite{Megow2014-3}. Electrostatic couplings
due to permanent charge distributions in the ground and the excited
state of the dye molecules result in the shift $\Delta{\cal E}_{m}^{{\rm (el)}}=\sum_{k}[J_{mk}(eg,ge)-J_{mk}(gg,gg)]$
of site--energy ${\cal E}_{m}$ (contributions due to solvent molecules
are incorporated in the $k$--sum \cite{Megow2010}).


The charge densities of the ground and excited states polarize the
environmental molecules, a polarization that reacts back on the considered
molecules and introduces the site energy shift $\Delta E_{m}^{({\rm pol)}}$.
A third type of site energy shift arises from the mutual Coulomb coupling
of transition densities and is termed dispersive site energy shift
$\Delta E_{m}^{({\rm disp)}}$ \cite{May2011,Renger2013}. Introducing
the dispersive change of the molecular energy referring to state $\varphi_{ma}$
as $\Delta E_{m}^{({\rm disp)}}$ we obtain (remember $a=g,e$) \cite{Megow2014-3}
\bea
\label{energy-shift}
\Delta E^{\rm (disp)}_{m a} = - \sum_k \sum_{f, f'} 
\frac{|J_{m k}(f' f, a \, g)|^2}{E_{m f a} + E_{k f' g}} \; .
\eea
The $f$ and $f'$ count all higher excited energy levels
($f,f'>e$), the $E_{mfa}$ and $E_{kf'g}$ are the transition energies
of molecule $m$ and $k$, respectively, and $J_{km}(f'f,a\, g)$
denotes the respective Coulomb matrix element. Such matrix elements
relate transitions in the considered molecule $m$ to transitions
in all other molecules labeled by $k$. According to the actual position
of molecule $m$ in the TDA the $k$--summation notices the concrete
conformation of dye molecules around molecule $m$. Consequently,
the overall site energy shift follows as $\Delta{\cal E}_{m}^{{\rm (disp)}}=\Delta E_{me}^{{\rm (disp)}}-\Delta E_{mg}^{{\rm (disp)}}$.

Hence, the site energies ${\cal E}_{m}$ are shifted in total by
\bea
\Delta {\cal E}_m=\Delta {\cal E}_m^{(\rm el)} + \Delta {\cal E}_m^{(\rm pol)}
+\Delta {\cal E}_m^{(\rm disp)}
\eea
For the present
C683 chromophore we find that $\Delta{\cal E}_{m}^{({\rm disp)}}>>\Delta{\cal E}_{m}^{({\rm el)}}$.
While $|\Delta{\cal E}_{m}^{({\rm el)}}|$ is in the order of 10 meV, $|\Delta{\cal E}_{m}^{({\rm disp)}}|$ is
in the order of over 500 meV.
This result is understood by noting the comparably similar charge
densities of the ground and excited electronic state (differential
dipole moment is about 1 D).
Since we expect
$\Delta{\cal E}_{m}^{({\rm el)}}$ in the same order of magnitude
as $\Delta{\cal E}_{m}^{({\rm pol)}}$, we get $\Delta{\cal E}_{m}^{({\rm disp)}}>>\Delta{\cal E}_{m}^{({\rm pol)}}$.
Therefore, we neglect $\Delta{\cal E}_{m}^{({\rm pol)}}$ in our calculations.


In order to determine the energy shifts of every individual molecule
within the aggregate the expression, Eq. \er{energy-shift}, has
to be transformed into a tractable form to compute its dependence
on the mutual position of the coupled molecules. Here, we follow Ref.
\cite{Megow2014-3}. Accordingly, in a first step $J_{mk}(ag,f'f)$
is replaced by an expression of two interacting extended dipoles with
charges $\pm q(fa)$ and $\pm q(f'g)$. Electronic structure computations
indicate that these dipoles are oriented along the elongated part
of the cyanine dye. Assuming uniform extension of all extended dipoles
we may write $\Delta{\cal E}_{m}^{{\rm (disp)}}=-Q\sum_{k}V_{mk}^{2}$
where $V_{mk}$ represents all geometrical factors, \textit{i.e.}
the Coulomb interaction of both extended dipoles but with unit charges.
Concrete values of the latter are included in the factor $Q=\sum_{f,f'}q^{2}(fe)q^{2}(f'g)[1/(E_{fe}+E_{f'g})-1/(E_{fg}+E_{f'g})]$,
where $E_{fe}$ ($E_{f^{\prime}e}$) and $E_{fg}$ ($E_{f^{\prime}g}$)
are the transition energies between the first excited and the ground
states of the isolated monomers, respectively, and the higher excited
states.


By further proceeding in line with Ref.~\cite{Megow2014-3} we do
not determine the $Q$--factor by a direct computation, which would
be not accurate enough, but by using experimental data on transition
energy shifts in non--polar solvents. If the solvent is described
by a dielectric continuum with refractive index $n$, the shift can
be approximately obtained as \cite{Bayliss1950,Renger2008},
\bea
\label{delta}
\Delta  {\cal E}_m^{\rm (disp)} = - {\cal F} \frac{n^2 - 1}{2n^2 + 1} \; ,
\eea
which describes the shift induced by
a polarizable environment. If a value for ${\cal F}$ is available
and if we know $n$ for a random arrangement of {\textbf C8S3} molecules we
can determine the site--energy shift. Noting previously published
spectra of a similar compound \cite{Wenyuan1985} we may deduce ${\cal F}\approx1.43$
eV, and a gas--phase transition energy follows as $E_{\text{gp}}=2.64$
eV. Moreover we take $n_{\text{TDA}}\approx1.78$ \cite{Shelkovnikov2012}.
According to \cite{Renger2008,Megow2014-3} we have to chose the long
wave--length limit of the refractive index resulting in $\Delta{\cal E}_{m}^{{\rm (disp)}}\approx-0.43$
eV. Identifying this value with $-Q\sum_{k}V_{mk}^{2}$ we may deduce
$Q$ after carrying out the $k$--summation. Since the experimental
value of $\Delta{\cal E}_{m}^{{\rm (disp)}}$ corresponds to a random
arrangement of the molecules (they have to form a structurless polarizable
continuum) we determine $Q=7.61$ eV\AA{}$^{2}${ }by introducing
an averaged expression $<\sum_{k}V_{mk}^{2}>$. It assumes random
mutual orientation of the molecules but at a mean density typical
for the TDA. 

This value of $Q$ is further used to calculate the dispersive site--energy
shifts of all molecules within the TDA. The used approximation leads
to a significant fluctuation of the individual energy shifts. This
is mainly caused by the extended dipole approximation, which is a
good approximation on average. However, individual couplings may vary
by up to 30 \%. Therefore, we average the shifts for the inner and
the outer cylinder separately.


The solvent effect on the dipersive site--energy shift $\Delta{\cal E}_{m}^{{\rm (disp)}}$
(Eq.~\ref{delta}) was neglected because of the small refractive
index of the solvent ($n\approx1.33$) and the comparably restricted
contact area between the solvent and the dye molecules. For comparison:
a dissolved monomer is shifted by about 0.3 eV (cf. Eq.~\ref{delta}).
The {\textbf C8S3} molecules within the tube however, obtain a smaller dispersive
shift, since they are mostly surrounded by other {\textbf C8S3} molecules (cf.~Fig.~\ref{fig2}).
Assuming that a molecule within a wall interacts with less than half
of the solvent molecules than a monomer in solution one obtains an
upper limit of below 0.15 eV for the solvent dispersive shift of {\textbf C8S3}
molecules within an aggregate. The dispersion shift due to solvent
interaction is somewhat larger for molecules in the outer cylinder,
since they have contact with more solvent molecules due to geometric
reasons. Assuming that this solvent interaction shift will be about
10 -- 15 \% larger for molecules in the outer cylinder 
one obtains a reduction of the mutual energy splitting of the two
J-aggregate bands by 0.02 eV (~6 nm) at most.

Having discussed the change of the site energies due to dispersion
of the environment we briefly comment on a related alternation of
the excitonic coupling, \textit{i.e.} we introduce screening to the
non--diagonal elements of Eq.~\ref{eq:Hagg}. Similar to the site--energy
shift it results from an environmental dispersion (see \textit{e.g.}
\cite{Adolphs2008}) which, in general, reduces the excitonic coupling
${\cal J}_{mn}$ between two molecules to $f_{mn}{\cal J}_{mn}$ \cite{Adolphs2008,Megow2014}.
The prefactor $f_{mn}$ can be approximated by a constant $1/\epsilon=1/n^{2}$
where $n$ is the optical refractive index \cite{Megow2014}. From
the dispersive shifts of the inner and outer cylinder we calculated
the respective refractive indices $n_{\text{out}}$ and $n_{\text{in}}$
for both cylinders. This was realized assuming the measured refractive
index $n=1.78$ being an average value $n=(N_{\text{out}}n_{\text{out}}+N_{\text{in}}n_{\text{in}})/(N_{\text{out}}+N_{\text{in}})$,
$N_{\text{out}}=468$ and $N_{\text{in}}=360$ being the numbers of
molecules in the outer and the inner cylinder, respectively. We obtain
$n=1.89$ for the inner cylinder and $n=1.69$ for the outer cylinder.
From the respective indices we yield screening factors for the coupling
of two molecules within a certain cylinder. We note that this site-dependent
screening factors, used for reasons of consistancy change our result
only slightly, which can be explained by the rather small contribution
of the excitonic couplings to the overall energy shifts.

For the coupling between two molecules located in different cylinders
we take $n=1.78$, getting $f_{mn}=0.35$ for the outer cylinder,
$f_{mn}=0.28$ for the inner cylinder and $f_{mn}=0.32$ for molecules
located in different cylinders, \textit{i.e.} a noticeable reduction
of the excitonic coupling. The somewhat larger screening factor for
the outer cylinder results in a slightly decreased energy distance
of the two J-aggregate peaks. The ${\cal J}_{mn}$ were calculated
via a coupling of transition partial charges (which have been somewhat
rescaled to reproduce the measured transition dipole moment \cite{Madjet2006}).

\subsection{Optical spectra}
After determining the site--energies and excitonic couplings which
define the exciton Hamiltonian $H_{{\rm ex}}$, Eq. \er{eq:Hagg},
in a proper way,
a respective diagonalization
results in exciton energies and wave functions. To account for a large
spatial delocalization of the exciton wave function, the TDA which
constitutes the Hamiltonian $H_{{\rm ex}}$ has been formed by five
fragments of 12.7 nm length used in the MD simulations (further increasing
of the number of fragments only induces a minor effect on the exciton
spectrum). Since after 7 ns of MD simulations all molecular positions
deviate from the ordered structure (see Fig.~\ref{fig1}) structural
disorder is accounted for. Moreover, the huge amount of molecules
which contribute to $H_{{\rm ex}}$ and thus to the exciton spectrum
introduce a self--averaging effect. Since it operates when computing
the absorption line--shape (sum of the squared excitonic transition
dipole moments times the energy conserving $\delta$--function) further
disorder averaging is not necessary. Homogeneous broadening was introduced
by replacing sharp transitions into the various exciton levels by
individual Lorentz--curves with a linewidth of 3 nm (FWHM).

\begin{figure}[!ht]
\includegraphics[width=0.45\textwidth]{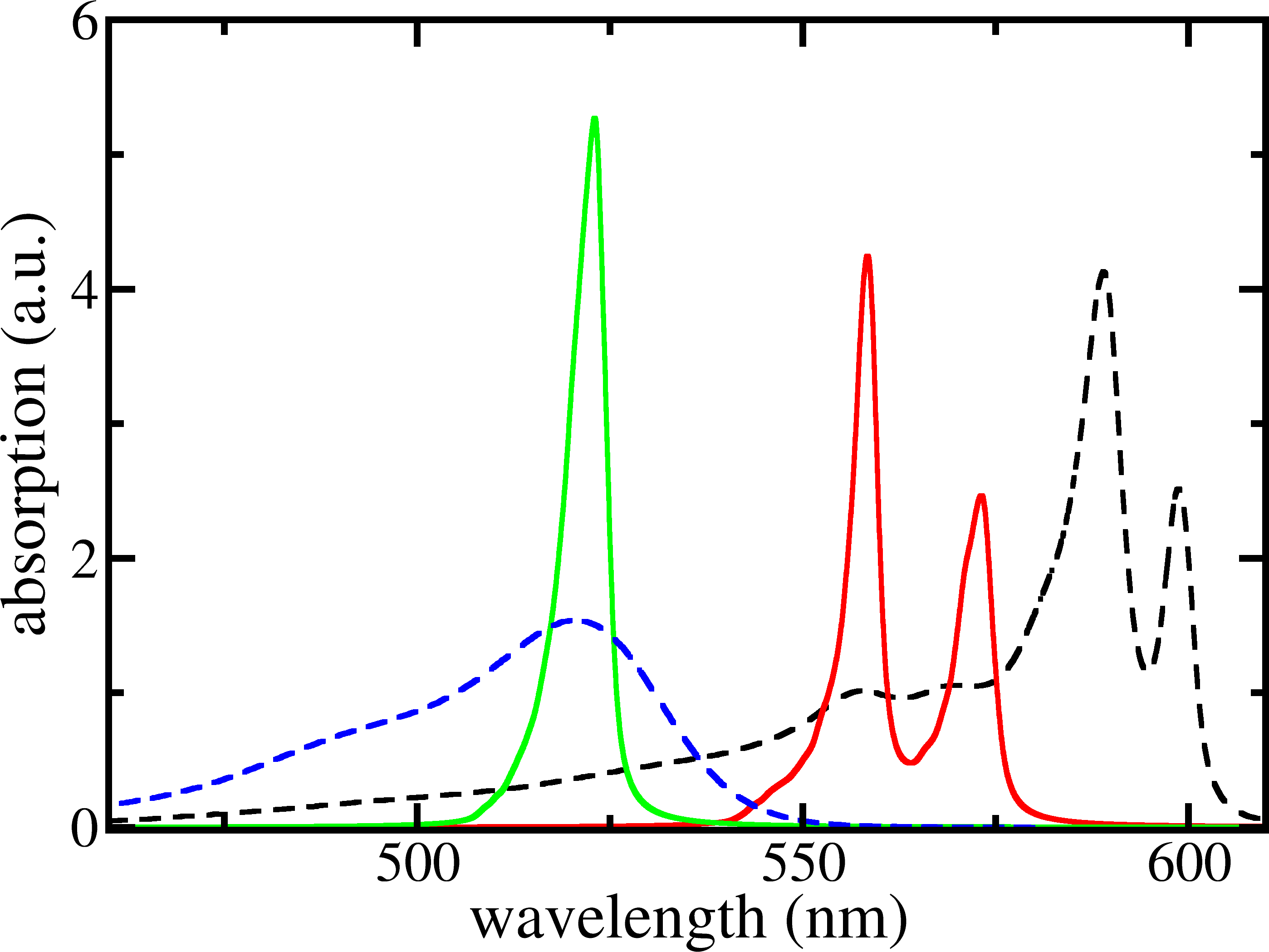}
\caption{{\footnotesize {Linear absorption spectra of a {\textbf C8S3} TDA and of respective
dye monomers. Black solid line: measured TDA absorption according
to \cite{Eisele2012}, blue dashed line: measured monomer absorption;
green line: calculated TDA absorption neglecting the site--dependent
dispersion; red solid line: completely calculated TDA absorption spectrum.
No free parameters used, except for the slip $\Delta$ in order to
build the structure.}}}

{\footnotesize \label{fig4} }
\end{figure}

Respective results together with measured data are presented in Fig.~\ref{fig4}.
If dispersive effects are neglected only a single TDA band results,
which is shifted to longer wavelengths compared with the monomer absorption.
Calculating the absorption spectra separately for the outer and the
inner TDA cylinder (yet neglecting dispersive effects) confirms almost
vanishing splitting between the two exciton bands. Including, however,
the site--dependent dispersion results in a line--shape that exhibits
two separate bands and agrees quite well with the measured curves.

The calculated red shift of the exciton bands is too small by 0.1
eV (about 30 nm) compared to the experimentally measured shift (see
Fig.~\ref{fig4}). This 0.1 eV offset can be related to the ignored
energy shift due to solvent dispersion but also to the uncertainty
when determining the linear factor ${\cal F}$ and the gas--phase
transition energy $E_{\text{gp}}$. This small unexplained shift has
to be compared to the total energy shift of more than 0.5 eV relative
to the $E_{\text{gp}}=$2.64 eV for which reason it can be accepted.

The treatment of solvent dispersion would also result in a slightly
smaller energy difference between the two J--aggregate peaks (0.02
eV at most).

\begin{figure}[!ht]
\includegraphics[width=0.45\textwidth]{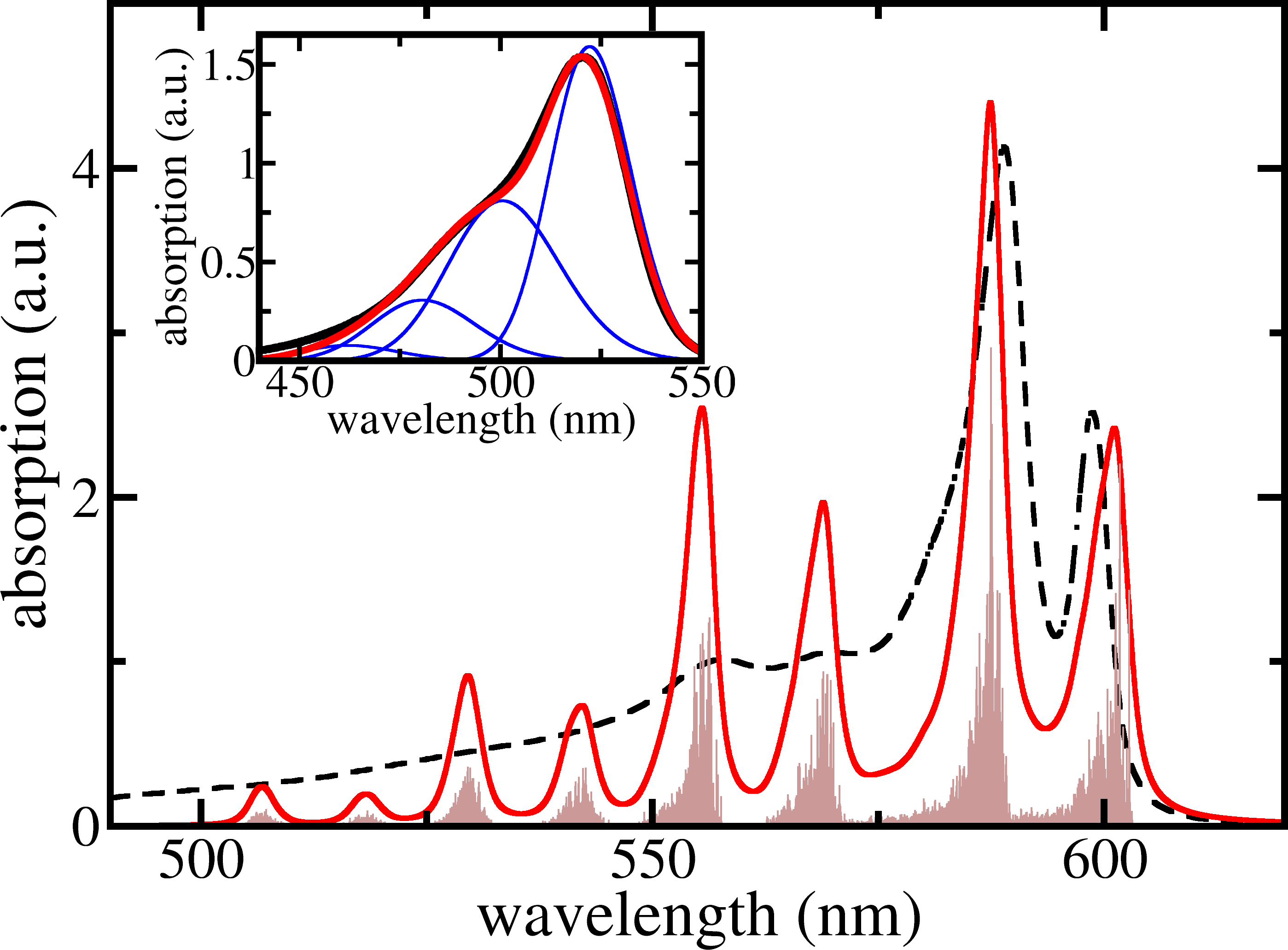}
\caption{{\footnotesize { Linear absorption spectra of a {\textbf C8S3} TDA including
a single representative intramolecular vibration per monomer (for
details see text). Orange filled area: stick spectrum, red curve:
$\approx$ 3 nm broadened stick spectrum, black dashed curve: measured
absorption (cf. Fig. \ref{fig4}). The offset of 0.1 eV due to solvent
dispersive shifts is included here. Insert: {\textbf C8S3} monomer absorption.
Red solid curve: computed absorption spectrum, black dashed curve:
measured spectrum, blue solid curves: single vibrational level contributions.
}}}

{\footnotesize \label{fig5} }
\end{figure}

Although the two long wavelength peaks of the TDA absorption are nicely
reproduced in their spectral position and height, there is a considerable
discrepancy around 550 nm. We briefly demonstrate that this discrepancy
may be overcome by including {\textbf C8S3} intra--molecular vibrations. Fig.
\ref{fig5} shows respective results where a single representative
and broadened vibrational mode per dye molecule has been used. Here,
the 0.1 eV offset seen in Fig.~\ref{fig4} and supposedly caused
by the neglection of solvent dispersive shifts is corrected. The coincidence
of the computed monomer spectrum with the measured one as shown in
the insert of Fig. \ref{fig5} justifies the used vibrational energy
of 0.1 eV, the Huang--Rhys factor of 0.74, and the line--broadening
of $\approx35$ nm (it differs somewhat between the $0\to0$ and $0\to\mu>0$
transitions).

To calculate the whole TDA exciton spectrum by including a single
vibrational mode per dye molecule we followed Ref. \cite{Spano2002}
and employed the so--called single particle approximation suggested
therein (cf. also \cite{Megow2014-3}). In this case, exciton states
are introduced as $\Phi_{\alpha}=\sum_{m,\mu}c_{\alpha}(m,\mu)\chi_{me\mu}\phi_{m}$,
where $\chi_{me\mu}$ is the total vibrational wave function with
$\mu=1...3$ vibrational excitations in the excited state of molecule
$m$. The expansion coefficients $c_{\alpha}(m,\mu)$ are obtained
via the diagonalization of the Hamiltonian, Eq. \er{eq:Hagg}, which,
however, has been extended on introducing vibrational contributions
in the site energies and excitonic couplings. Now, vibrational satellites
appear in the wave--length region around 550 nm which corresponds
to absorption bands observed experimentally. The remaining differences
between the calculated spectrum and the experimental data may be caused
by the restriction to a single vibrational mode per {\textbf C8S3} monomer.
At this point we cannot exclude that other configurations are possible
that lead to additional exciton bands and thus would reproduce more
details of the spectra without involving vibrational modes. However,
we can state that vibrational modes seem to strongly contribute to
the spectra, what is due to the rather large Huang--Rhys factor. We
can state further that our model is consistent regarding cryo--TEM
figures, optical spectra, electronic structure calculations and MD
simulations. Considering Fig.~\ref{fig4}, the only free parameter
is the slip $\Delta$ between the two molecules in the unit cell,
which was chosen to reproduce the spectra.


\section{Discussion}

A novel structure model for a {\textbf C8S3} TDA together with
site--dependent dispersive effects leads to a conclusive interpretation
of the general features of the respective linear absorption spectra.
Molecules in the inner and outer cylinder of the TDA interact with
a different environment, and accordingly their dispersive shifts differ.
For the considered {\textbf C8S3} TDA the excitonic coupling is particularly
small due to screening and, thus, the observed absorption line splitting
is mostly determined by environmental dispersive effects. This splitting
would vanish if the dyes formed an uncurved double--walled layer.
Accordingly, we consider a long--wavelength absorption line splitting
in TDAs a general feature, due to a curvature--induced site--dependent
dispersive effect.
Therefore we conclude that the site--dependence of dispersive energies
is of key--importance to understand the optical properties of nanostructured
molecular dye systems, such as nano--tubular light harvesting dye aggregates.
We strongly suspect that this reasoning holds for nanoscaled molecular aggregates in general. \\
\section{Methods}
\subsection{MD-simulations}
The MD simulations have been carried out
using NAMD \citep{Phillips2005} together with the AMBER force field
\citep{Case2004} as well as the GAFF parameter sets \citep{Wang2004-2}.
The MD-simulations are carried out in a periodic box, the electrostatic
interactions were computed using the particle mesh Ewald method \cite{Darden1993}.
The temperature was heated to 300 K and 7 ns of MD simulation with
a time-step of 0.5 fs was carried out.

\subsection{DFT-calculations}
The geometry of the monomeric cyanine dye
has been optimized at the DFT level, the absorption energy for the
first excited state has been calculated in the frame of TDDFT employing
the long-range corrected hybrid CAM-B3LYP functional \cite{Yanai2004}
and the TZVP atomic orbital basis set \cite{Weigend2005} for all
atoms. Atomically centered (static) partial charges (electronic ground and
first excited state) and transition charges (only for heavy atoms)
have been fitted to the electrostatic potential of the respective
charge/ transition charge density by using the CHELPG procedure \cite{Breneman1990}.

The static partial charges have been rescaled with respect to
the differential dipole moment $\Delta{\bf d}$ that was measured from hole burning experiments for similar cyanine dyes
($0.7D<|\Delta{\bf d}|<1.1D$)\cite{Chowdhury2001}.
We chose $\Delta{\bf d}=1D$ and we note that other values within
the range will change the absorption spectrum only slightly.


\textit{Acknowledgments:} 
Financial support by the \textit{Deutsche Forschungsgemeinschaft}
through Sfb 951 and through project ME 4215/2-1 (J.M.) and by the
\textit{Austrian Science Fund} (FWF) through project P 24774-N27 (T.R.)
are gratefully acknowledged.




\end{document}